\documentclass[journal]{IEEEtran}
\usepackage{graphicx}
\usepackage[caption=false,font=footnotesize]{subfig}  
\usepackage{newtxtext,newtxmath}
\usepackage{textcomp}   
\usepackage{xcolor}
\usepackage{url}
\usepackage{cite}
\usepackage{amsmath}
\usepackage{booktabs}
\usepackage{multirow}
\usepackage{threeparttable}
\usepackage{array}
\usepackage{siunitx}
\usepackage{hyperref}
\begin{document}

\title{On the Potential of Electrified Supply Chains\\to Provide Long Duration Demand Flexibility}

\author{Rina~Davila~Severiano,~\IEEEmembership{Student Member,~IEEE,}
        Constance~Crozier,~\IEEEmembership{Member,~IEEE,}
        and~Mark~O'Mailley,~\IEEEmembership{Fellow,~IEEE}
\thanks{R. Davila Severiano and C. Crozier are with the H Milton Stewart School of Industrial and Systems Engineering, Georgia Institute of Technology, USA. M. O'Mailley is with the Department of Electrical and Electronic Engineering at Imperial College London}}

\maketitle

\begin{abstract}
Demand flexibility can offset some of the variability introduced on the supply-side by variable renewable generation. However, most efforts (e.g. control of residential vehicle charging) focus on short durations -- typically on the scale of minutes to hours. This paper investigates whether a fully electrified supply chain (transport and manufacturing) could provide demand flexibility over longer durations, exploiting the latency that typically exists between the processing of raw material to the delivery of finished product. Using a case study of the cement industry along the East Coast of the United States, we demonstrate that electrified supply chains could shift gigawatt-hours (GWh) of electricity demand for durations of more than a week, largely following wind power variability. Furthermore, we show that this occurs using low levels of carbon taxing (below \$50/tn), at which battery storage is not economically viable. A sensitivity analysis shows potential to provide flexibility in all considered cost scenarios, although where the flexibility comes from can change (e.g. transport vs manufacturing). We show that today's cost of electrified heavy goods vehicles are the most significant parameter -- with substantially lower costs yielding a more demand-flexible supply chain. 
\end{abstract}


%

\IEEEpeerreviewmaketitle

\section*{Nomenclature}
\begin{tabular}{ll}
& \textbf{Decision variables}\\[.1cm]
$m_{s}$ & The amount of process $s$ taking place\\
$p$ & The total consumed power (kW)\\
$y$ & The number of loaded trucks\\
$y'$ & The number of empty trucks\\
$z$ & The total non-renewable power (kW)\\
$r_{curt}$ & The curtailed renewable power (kW)\\
$C$ & The cumulative charge of vehicles (kWh)\\
$F$ & The amount of raw material available (kg)\\
$M_{s}$ & The amount of process $s$ that can occur\\
$W$ & The capacity of the storage facility (kg)\\
$X$ & The amount of product stored (kg)\\
$Y$ & The number of stationary trucks\\
$\hat{Y}$ & The total number of trucks in the fleet\\[.1cm]
&\textbf{Parameters}\\[.1cm]
$f$ & The amount of raw material arriving (kg)\\
$q$ & Product injection: supply (+), demand (-) (kg)\\
$r$ & The renewable power available (kW)\\
$E$ & The amount of energy required to drive (kWh)\\
$B^{max}$ & The amount of installed battery storage (kWh)\\
$N_t$ & Number of time-steps in the planning horizon\\
\end{tabular}

\begin{tabular}{ll}
$T_{batt}$ & Truck battery capacity (kWh)\\
$T_{load}$ & Truck load carrying capacity (kg)\\
$T_{w}$ & Truck weight unloaded (kg)\\
$k^{batt}$ & Levelized battery cost (\$/kWh)\\
$k^{equip}$ & Levelized equipment cost (\$/unit-hr)\\
$k^{power}$ & Penalty for non-renewable power (\$/kW)\\
$k^{truck}$ & Levelized cost of a truck (\$/hr)\\
$k_i^{store}$ & Levelized cost of product storage at $i$ (\$/kg-hr)\\
$\mu$ & The number of hours for a truck to fully charge\\
$\theta$ & The ratio of an unloaded vs. loaded truck weight\\
$\Delta_t$ & The size of a single time step (hrs)\\
$\tau_{ij}$ & Number of time-steps to travel from $i$ to $j$\\
$\tau_{s}$ & Number of time-steps which process $s$ runs for\\
$\alpha_s$ & Electricity consumption of process $s$ (kW/unit)\\
$\beta_s$ & Production rate of process $s$ (kg/unit)\\
$\gamma_s$ & Raw material consumption of process $s$ (kg/unit)\\[.1cm]
& \textbf{Notation}\\[.1cm]
$^{(t)}$ & At time step $t$\\
$_i$ & At location $i$\\
$_{i,j}$ & Along the path from $i$ to $j$\\
$_s$ & Relating to process $s$
\end{tabular}

\section{Introduction}
%
%
%
%

\IEEEPARstart{T}{his} paper investigates the potential of supply chains with electrified transport and manufacturing to provide long duration load shifting to the power system. 

Electrification of both transportation and manufacturing is being explored as a mechanism to reduce industrial and transportation emissions. Using electricity instead of carbon-based fuels can significantly reduce emissions, as there are developed technologies which can produce low-carbon renewable electricity, such as wind and solar power. 

A major challenge of integrating wind and solar power into the power grid is their inherent variability. Battery energy storage systems are a common method of addressing fluctuations, and have been shown to be effective in increasing the use of renewable energy~\cite{7355371}. However, a major challenge is their cost, meaning the economic viability of the systems relies on frequent use~\cite{HU2022119512,MWASILU2014501,10286401}. Therefore, while battery storage may be suitable for tackling high-frequency variation, they are poorly-suited to tackling longer duration variation (e.g., several days of low wind output). There are other forms of energy storage that may be better suited for longer durations, but they largely have low levels of technology development~\cite{WANG2022104812,JENKINS20212241}.

An alternative approach for managing fluctuations in renewables is to adjust energy demand to match these fluctuations. Demand response aims to adjust the consumption of electricity in response to a signal such as price~\cite{10938867}. Electric vehicle (EV) charging is a good candidate for demand response; often, vehicles remain parked at chargers for many hours longer than necessary, so charging can be delayed without impacting owners. Previous work has demonstrated the grid potential for both uni-directional smart charging~\cite{crozier2020opportunity,DAS2020109618,10695045} and bi-directional vehicle-to-grid services (where the vehicle may also discharge back to the grid)~\cite{crozier2020case}. The vast majority of this work focuses on privately owned residential vehicles, which have hard constraints on use (e.g. that the vehicle needs to be charged in time for the morning commute). Some more recent work focuses on fleet operation, such as buses~\cite{MANZOLLI2022124252,9266099, PURNELL2022118272}, taxis~\cite{YU2024122323}, or delivery fleets~\cite{MOHAMMADIAN2025125036, HILL2012221,WANG2024123407,10741677}. In these cases, trip constraints are not tied to a single vehicle, meaning some flexibility in vehicle scheduling can be exploited alongside the charging flexibility. This is often referred to as coupling between the power and transport systems, where we extend vehicle routing problems to include charging and energy constraints. Approaches also vary on their objective; charging as fast as possible~\cite{6730964}, only at their home base~\cite{ALHANAHI2022100182}, minimizing energy consumption~\cite{WOO2024103644}, or minimizing charging costs~\cite{MOHAMMADIAN2025125036}. In general, EV charging is considered only to be able to provide demand flexibility over shorter durations (hours)~\cite{wevj10010014,NOLAN20151}. 

Recent work has explored the potential of manufacturing industries to provide longer duration load shifting, in cases where there is enough flexibility in the manufacturing scheduling~\cite{YANG2020106846}. Several papers have explored single facilities adapting their electricity use in response to price signals. Y. Wang et al. \cite{WANG2013233} demonstrate how time-of-use electricity pricing, can be incorporated into manufacturing systems to generate a more efficient production schedule. In \cite{LU2021116291}, the authors explore how steel powder manufacturing facilities utilize real-time pricing to manage their electricity demand response. Similarly, \cite{7938652} examines a problem involving decision-making for electricity demand in response to real-time pricing. These works, and others, consider the operation of a single geographic location without including a transportation network.

\begin{figure}[htbp]
    \centering
    \includegraphics[width=\columnwidth,trim={.9cm .6cm 0cm 0},clip]{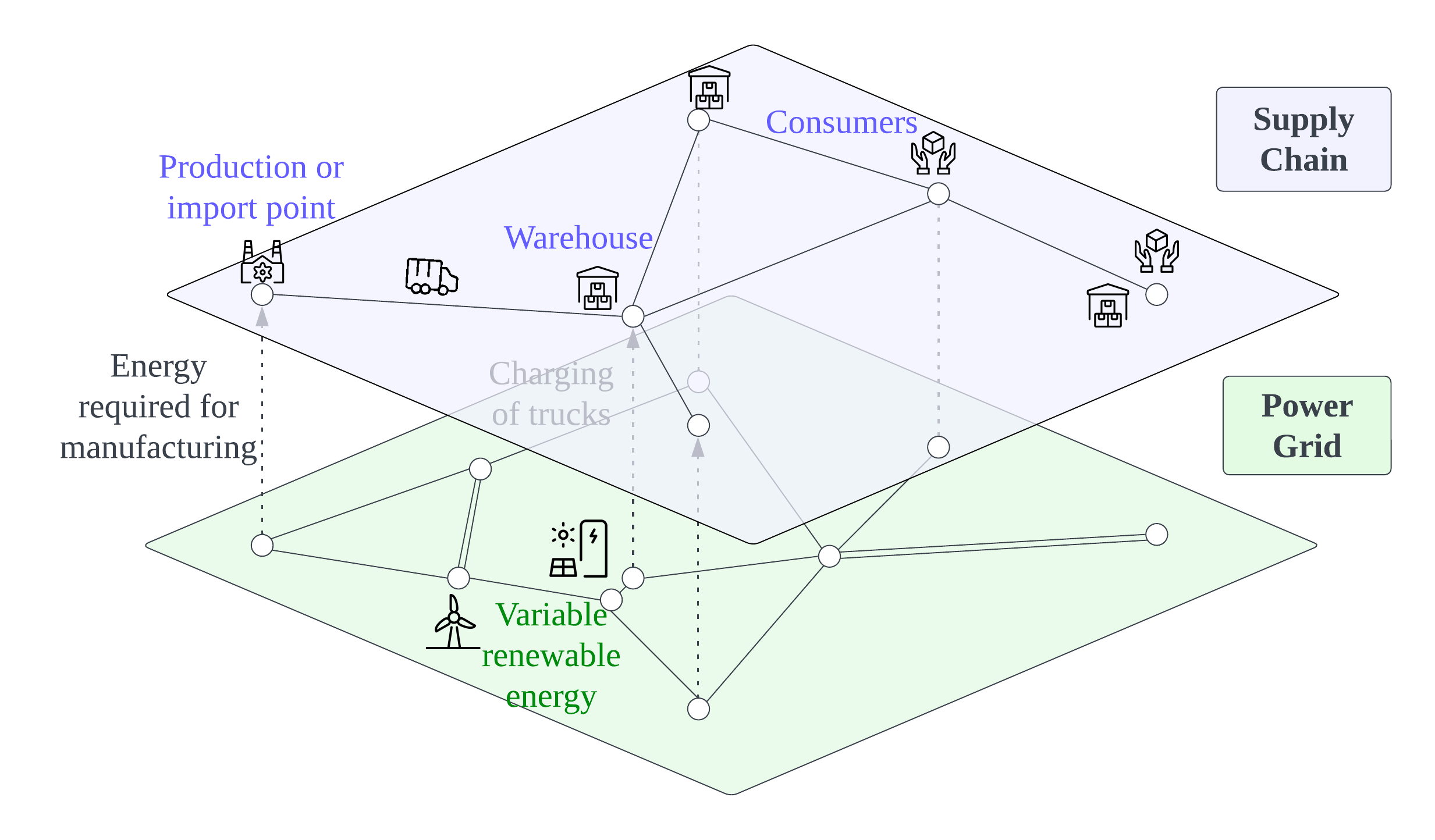}
    \caption{A visualization of the coupled supply chain and power network operation problem, illustrating how renewable energy generated in the power system is used within the supply chain to manufacture finished products. These products are stored in a warehouse and transported to consumers using electric trucks, which are charged along the route.}
    \label{fig:enter-label}
    \vspace{-.3cm}
\end{figure}

While considering a transportation fleet or a network of manufacturing facilities in isolation, it is not possible to quantify the demand flexibility potential; delaying manufacturing is only possible providing finished products can still be transported to their destination in time. In this paper, we consider the centralized operation of a fully electrified supply chain (visualized in Fig. \ref{fig:enter-label}) to investigate whether demand flexibility could be economically provided over longer durations (days--weeks).

The contributions of this paper can be summarized as follows. First, we formulate a problem for the centralized operation of a supply chain with electrified transport and manufacturing. Second, we explore the ability of a single product supply chain to economically perform demand flexibility over longer durations (days--weeks). Finally, we investigate which qualities make a supply chain most suitable for providing demand flexibility -- using a broad sensitivity analysis to understand the most significant cost and demand parameters. 

\section{Centralized Planning Formulation}

\begin{figure}[htbp]
    \centering
    \includegraphics[width=\columnwidth,height=9cm,keepaspectratio]{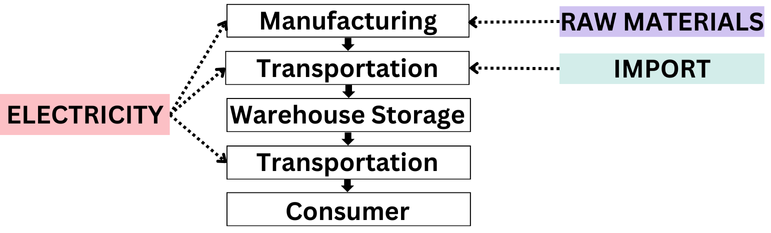}
    \caption{Diagram illustrating the electrified supply chain process for finished products. The key stages include manufacturing (using raw materials), transportation, warehouse storage, and consumer delivery of the finished product, along with imports of additional finished products.}
    \vspace{-.4cm}
    \label{fig:flow_diagram}
\end{figure}

In this section, we provide the details of our proposed model, which leverages supply chain flexibility and considers the centralized operation of a supply chain with electrified manufacturing and freight transportation. In Fig. \ref{fig:flow_diagram} we visualize the flow of finished products in the supply chain. Here we break down the problem into three main sections: electrified manufacturing, electrified transport, and renewable generation.

\subsection{Electrified Manufacturing}
We define a number of manufacturing processes $s$. Each process is assumed to last $\tau_s$ time-steps during which it uniformly consumes $\alpha_s$ kW of power per unit of production. Each unit of the process requires $\gamma_s$ kg of raw material $F_{s}$ and produces $\beta_s$ kg of the finished product.

It is necessary to consider some exogenous movements of both the finished product and raw materials. Here we assume that the supply of raw materials, imports of additional finished products, and demand for the finished product are not controllable. Note that this means the costs and revenues of imports and sales are not included in the objective, as they are assumed to be fixed. We define a parameter $q_i^{(t)}$ to be the exogenous supply of products at location $i$ and time $t$; positive values reflect imported finished products from another manufacturer (e.g., international imports), and negative values represent the demand for the finished product (which we assume must be met). We also define the variable $f_{i}^{(t)}$ to be kg of raw material arriving at location $i$ and time $t$. 

\subsection{Electrified Transport}
We assume a fleet of $\hat{Y}$ electric trucks with a maximum battery capacity of $T_{batt}$. We consider the possible paths $ij$ to be the set of origin-destination pairs that a truck could reach with a single charge -- in other words, where the required energy $E_{ij}\leq T_{batt}$. At any given time, the number of trucks stationary at each location $i$ is given by $Y_{i}^{(t)}$, and these are assumed to be available for charging/discharging. Note that we do not consider exporting to the power grid (known as vehicle-to-grid) but do consider possible discharging to the manufacturing facility (known as vehicle-to-building). The variable $y_{ij}^{(t)}$ describes the number of fully loaded trucks, carrying finished product, dispatched from location $i$ to location $j$ at time $t$, and $y_{ij}^{'(t)}$ is the number of dispatched empty trucks. It is necessary to distinguish between fully loaded and empty trucks because of their varied impact on the movement of finished products and energy consumption.

\subsection{Renewable Generation}

We consider that $r^{(t)}_i$ kW of renewable power is available at time $t$ and location $i$. To incentivize the use of renewables, we include a cost term in the objective that penalizes any non-renewable power used. Effectively, this introduces a price differential between renewable and non-renewable power. Later in the paper, we interpret this differential as a carbon tax. We consider two levels: one equivalent to a \$50/ton carbon tax and another equivalent to \$250/ton. The former reflects a relatively conservative estimate, while the latter remains within the range currently being considered by some countries \cite{2312030121}. To convert the carbon tax into a price penalty, we use an emission factor of $0.389\frac {\text{kg CO2}}{\text{kWh}}$, assuming natural gas is the marginal generator. This means that a \$50/tn carbon tax is equivalent to a cost differential of \$19.45/MWh between renewable and carbon-based power. 

\subsection{Optimization Formulation}
We consider the operation of the coupled system over a planning horizon of $N_t$ discrete time-steps of size $\Delta_t$ hours. Our cost-minimizing objective is as follows:
\begin{align}
\text{min}\quad \Delta_t N_t f_{capex} + f_{opex}\,,
\end{align}
where $f_{capex}$ represents the capital costs levelized over the component lifetimes (\$/hour), and $f_{opex}$ represents the operational costs over the considered time-horizon ($\$$). 
The capital costs include the costs of purchasing trucks, a warehouse, and manufacturing equipment, and can be calculated as follows:
\begin{align}
f_{capex} &= k^{truck}\hat{Y} + \sum_i (k_i^{store} W_i + \sum_s k_s^{equip} M_{s,i})
\end{align}
Where the variables $k^{truck},k^{store},k^{equip}$ give the capital costs per unit of the trucks, warehouse, and equipment, respectively. $W_i$ defines the maximum capacity of the warehouse at location $i$, and $M_{s,i}$ gives the maximum rate of production of process $s$ in location $i$. The operation costs considered are related to the consumption of electricity:
\begin{align}\label{eq:opex}
f_{opex} &= k^{power}  \sum_t z^{(t)} \Delta_t +  \varepsilon\sum_t\sum_i p_{i}^{(t)}\,
\end{align}
where $k^{power}$ is the penalty applied to non-renewable power, and $\varepsilon$ represents a small base cost applied to all electricity consumed. The variable $z^{(t)}$ represents the non-renewable power used at time $t$, while $p_i^{(t)}$ is the total power (both non-renewable and renewable) at time $t$ and location $i$.

We include variables for each of the system states: the accumulated charge $C_i^{(t)}$, the number of stationary trucks $Y_i^{(t)}$, the stored finished product $X_i^{(t)}$, and the stored raw materials $F_i^{(t)}$. We define each of these states for location $i$ and time $t$, with each evolving according to system dynamics over time. The following constraints enforce state updates for all locations $i$ and times $t$:

\begin{subequations}\label{eq:updates}
\begin{align}
\begin{split}
 C_{i}^{(t)}- C_{i}^{(t-1)}  =& \Delta_t \Big(p^{(t)}_i -  \sum_s \sum_{\tau=0}^{\tau_s}\alpha_s m_{s,i}^{(t-\tau)}\Big) \\ & - \sum_j E_{i,j}(y_{i,j}^{(t)} + \theta {y'_{ij}}^{(t)})    
\end{split}
\label{eq:c_update}\\
\begin{split}
Y_{i}^{(t)} - Y_{i}^{(t-1)} =& \sum_j \Big(y_{ij}^{(t)} - {y'_{ij}}^{(t)} + y_{ji}^{(t-\tau_{ij})}  \\&+  {y'_{ij}}^{(t-\tau_{ij})} \Big)
\end{split}\label{eq:y_update}\\
\begin{split}
    X_{i}^{(t)} - X_{i}^{(t-1)} =& q_{i}^{(t)} + T_{load} \sum_j (y_{ji}^{(t-\tau_{ij})}-y_{ij}^{(t)})\\ &+ \sum_s \beta_s m_{s,i}^{(t-\tau_s)}
\end{split}\label{eq:x_update}\\
F_{i}^{(t)} - F_{i}^{(t-1)} =& f_{i}^{(t)} - \sum_s \gamma_s m_{s,i}^{(t)} \label{eq:f_update}
\end{align}
\end{subequations}
Equation \eqref{eq:c_update} updates the accumulated charge at each location $C_{i}^{(t)}$. The first term describes the energy from grid charging, the second term represents the energy used by departing trucks (both load-carrying and empty), and the final term accounts for energy used in manufacturing processes. The parameter $E_{ij}$ is the amount of energy required to drive from location $i$ to location $j$. The variable $m_{s,i}^t$ is the amount of process $s$ taking place in location $i$ at time $t$. Note that the expression $\theta=\frac{T_w }{T_{w}+ T_{load}}$ represents the ratio of the weight of an unloaded $T_{w}$ vs. loaded truck $T_{load}$ -- it is assumed that the energy consumption of the truck varies linearly with weight~\cite{crozier2018numerical}.
Equation \eqref{eq:y_update} updates the number of stationary trucks $Y_i^{(t)}$, where the first two terms account for departing full and empty trucks and the latter two terms account for arriving trucks (note the delayed time index to account for travel time $\tau_{ij}$). Equation \eqref{eq:x_update} updates the stored product $X_i^{(t)}$. The first term accounts for arriving imports, the second term describes the departing and arriving finished product from the trucks, and the final term describes the finished products produced from the manufacturing process. Finally, equation \eqref{eq:f_update} updates the stored raw material $F_{i}^t$, with the terms representing the imported raw material $f_{i}^{(t)}$ and the raw material required for each process $s$.


Additionally, the following constraints are required in order to include relational limits between variables:
\begin{subequations}
\begin{align}
\begin{split}
 C_{i}^{(t)}- C_{i}^{(t-1)} \leq & \frac{1}{\mu}\Delta_t T_{batt}Y_i^{(t-1)} 
\end{split}
\label{eq:c_capacity}\\
\begin{split}
    C_{i}^{(t)} &\leq Y_i^{(t)} T_{batt}  \quad \forall i,t
\end{split} \label{eq:c_max}\\
\begin{split} W_i &\geq  X_{i}^{(t)} + \sum_s F_{s,i}^{(t)}\quad \forall i,t 
\end{split}\label{eq:x_max}
\end{align}
\end{subequations}
\begin{subequations}
\begin{align}
\begin{split}
    \hat{Y} &\geq  \sum_i Y_i^{(t)} + \sum_{ij}\sum_{\tau=0}^{\tau_{ij}} y_{ij}^{(t-\tau)} \quad \forall t 
\end{split} \label{eq:y_max}\\
\begin{split}
    M_{s,i} &\geq  \sum_{\tau=0}^{\tau_s} m_{s,i}^{(t-\tau)}\quad \forall i,s,t
\end{split}\label{eq:m_max}\,
\end{align}
\end{subequations}
Equation \eqref{eq:c_capacity} ensures that the accumulative charge $C_{i}^{(t)}$ of trucks at location $i$ is bounded according to the charger capacity, where the parameter $\mu $ is the number of hours for a truck to charge fully. 
Equation \eqref{eq:c_max} ensures that the total accumulated charge at each location is less than the total battery capacity of the trucks at that location. Implicitly we assume that the charging capacity is unlimited (i.e. that there is no limit to how many trucks can charge simultaneously at each location) and that charging is lossless. These additional features could be added relatively easily, but here we neglect them for computational tractability -- noting that our results will present an upper bound compared to the lossy and congested case. 

Equation \eqref{eq:x_max} ensures that the total accumulated finished product and raw material fit within the warehouse at each location. Equation \eqref{eq:y_max} ensures that the total stationary and moving trucks are within the total truck fleet size. Note that this expression would naturally be an equality (the trucks should all be accounted for). However, given that the objective includes $\hat{Y}$, the inequality should always be active -- we leave the inequality for computational simplicity. Finally, equation \eqref{eq:m_max} ensures that there is sufficient equipment available at location $i$ to complete the chosen manufacturing processes. 

Finally, we include boundary constraints on the stored product and accumulated charge:

\begin{subequations}
\begin{align}
   X_{i}^{(1)} &= X_{i}^{(N_t)}, \quad \forall i  \label{eq:product} \\
   C_{i}^{(1)} &= C_{i}^{(N_t)}, \quad \forall i  \label{eq:charge}
\end{align}
\end{subequations}

Constraints \eqref{eq:product} and \eqref{eq:charge} are used to prevent the model from exploiting boundary conditions, ensuring that there is no net gain or loss of stored product or energy over the planning horizon. Additionally, we need to introduce positivity requirements for all decision variables:
\begin{align}
    z^{(t)}, y_{ij}^{(t)}, y_{ij}^{'(t)} m_{s,i}^{(t)}, F_{i}^{(t)}, X_i^{(t)}, W_i^{(t)}, Y_i^{(t)}, \hat{Y}, C_i^{(t)} &\geq 0 \,
\end{align}
Coupled with the state update equations~\eqref{eq:c_update}--\eqref{eq:f_update}, these ensure that sufficient energy, materials, vehicles, and warehouse are available for operations. For example, by ensuring $C_i^{(t)} \geq 0$ we ensure that the trucks have all charged sufficiently before departure and that enough power is procured for the manufacturing processes. 


Finally, we need to introduce a power-balancing model. In this paper, we include a \textit{copper plate} model -- meaning that the distribution of power is not considered, we only consider the balance of total supply and demand. This can be achieved using the following:
\begin{align}
    \sum_i p^{(t)}_i &\leq \sum_i r_i^{(t)} + z^{(t)} \quad \forall t\,.
\end{align}
This equation ensures that the total power drawn from the grid across all locations is equal to the total renewable power plus the total non-renewable power. If a spatial power system model were to be considered, this constraint would be replaced with some form of the power flow equations. 

\subsection{Battery energy storage equivalent}

Finally, we introduce a Battery Energy Storage System (BESS) model to compare the cost-effectivity of utilizing supply chain flexibility compared to grid scale battery storage. 
In the BESS model, we consider whether a central operator would install batteries at charging and/or manufacturing facilities given the costs on non-renewable defined in \eqref{eq:opex}. For the BESS model, we do not consider flexible operation of the supply chain. Instead, the truck and manufacturing schedule are derived from the cost-minimizing operation of the coupled system, where no carbon tax is imposed on non-renewable power usage ($k^{power}=0$).

The objective of the BESS model is to minimize the total cost, which includes the cost of installed batteries and the power penalty on non-renewable power. The optimization formulation is as follows:

\begin{subequations}
    \begin{align}
        \min \quad k^{batt} B^{max} + k^{power} \sum_t z^{(t)} + \varepsilon\sum_t p^{(t)}\,,   \end{align}
\end{subequations}
where $B^{max}$ is the amount of installed battery storage and $k^{batt}$ is the levelized capital cost of the battery (scaled to the planning horizon as a percentage of the expected battery lifetime). The operational costs for electricity purchases are the same as for the supply chain flexibility model, defined in \eqref{eq:opex}.

For the BESS model, we consider three main constraints:
\begin{subequations}
    \begin{align}
        B^{(t)} - B^{(t-1)} &= -p^{(t)} + r^{(t)} + z^{(t)} - r_{curt}^t \quad \forall t \label{eq:1st} \\
        B^{(t)} &\leq B^{max} \quad \forall t \label{eq:2nd} \\
        B^{(0)} &= B^{(N_t)} \label{eq:3rd}
    \end{align}
\end{subequations}

Equation \eqref{eq:1st}, ensures that \( B^{(t)} \), the energy stored in the battery at time \( t \), is updated based on \( p^{(t)} \), the total power consumed at time \( t \); \( r^{(t)} \), the renewable power available at time \( t \); \( z^{(t)} \), the non-renewable power generated at time \( t \); and \( r_{\text{curt}}^{(t)} \), the curtailed renewable power at time \( t \). Equation \eqref{eq:2nd} make sure that the stored energy \( B^{(t)} \) does not exceed the battery's maximum capacity \( M_{\text{batt}} \). Finally, equation \eqref{eq:3rd} requires the amount of energy stored initially, $t = 0$,  to be the same as the energy stored at the end of the planning period, $t = N_t$. Finally, positivity requirements need to be introduced for all the decision variables:
\begin{align}
B^{max}, B^{(t)}, z^{(t)}, r_{curt}^t  \geq 0 \,.
\end{align}

\section{Case Study}

In this section, we introduce the details and modeling assumptions behind our electrified supply chain case study. For this analysis, we sought a product whose manufacturing process is a significant energy consumer. We chose cement manufacturing for its dominant heating energy consumption -- requiring a blended mixture to be heated to 1400$^\circ C$ in a kiln.  Furthermore, cement is one of the largest contributors to $\text{CO}_2$ emissions, highlighting the critical importance of electrifying its production process and using renewable energy sources for power. In this section, we describe the setup of our case study, including the input parameters and geographic network considered.

\subsection{Case Study Parameters}

For cement manufacturing, we assume a single-stage process with the parameters shown in Table \ref{t: Process}. We have included details of the derivation of these numbers in the Appendix. State-level cement demand was calculated using data from the United States Geological Survey's (USGS) Mineral Industry Surveys, which provide monthly figures for Blended and Portland cement shipments in metric tons \cite{usgs2024cement}. An average month was created using data from January to June 2024.

For the electric heavy goods vehicles, we assume that the trucks require two hours to fully charge, implying $\mu = 2$. Given a total battery capacity of 900 kWh, this corresponds to a charging power of 450 kW. These parameters align with the 2024 Tesla Semi-Truck \cite{ali2025tesla} and the current fastest DC chargers. Using the same vehicle model, we assume trucks are priced at \$150,000 per truck, with a 30-year lifetime, and a total load carrying capacity of 20,000 kg. For the grid-scale battery storage comparison, we assume a cost of \$400 per kWh and a lifetime of five years. 

\begin{table}[h!]
\centering
\caption{Assumed parameters for the cement manufacturing process, derived in the Appendix.}
\begin{tabular}{l|l|l}
\toprule
\textbf{Parameter} & \textbf{Value} & \textbf{Unit} \\
\midrule
Process Energy Consumption (\(\alpha\)) & 56,667 & kWh \\
Production Rate (\(\beta\)) & 150,000 & kg/unit \\
Rate of raw material Consumption (\(\gamma\)) & 150,000 & kg/unit \\
Process Duration & 1 & hours \\
\bottomrule
\end{tabular}
\label{t: Process}
\end{table}
 
\begin{figure}[htbp]
    \centering
    \includegraphics[width=\columnwidth,trim={2.9cm 3.cm 3.5cm 4.3cm},clip]{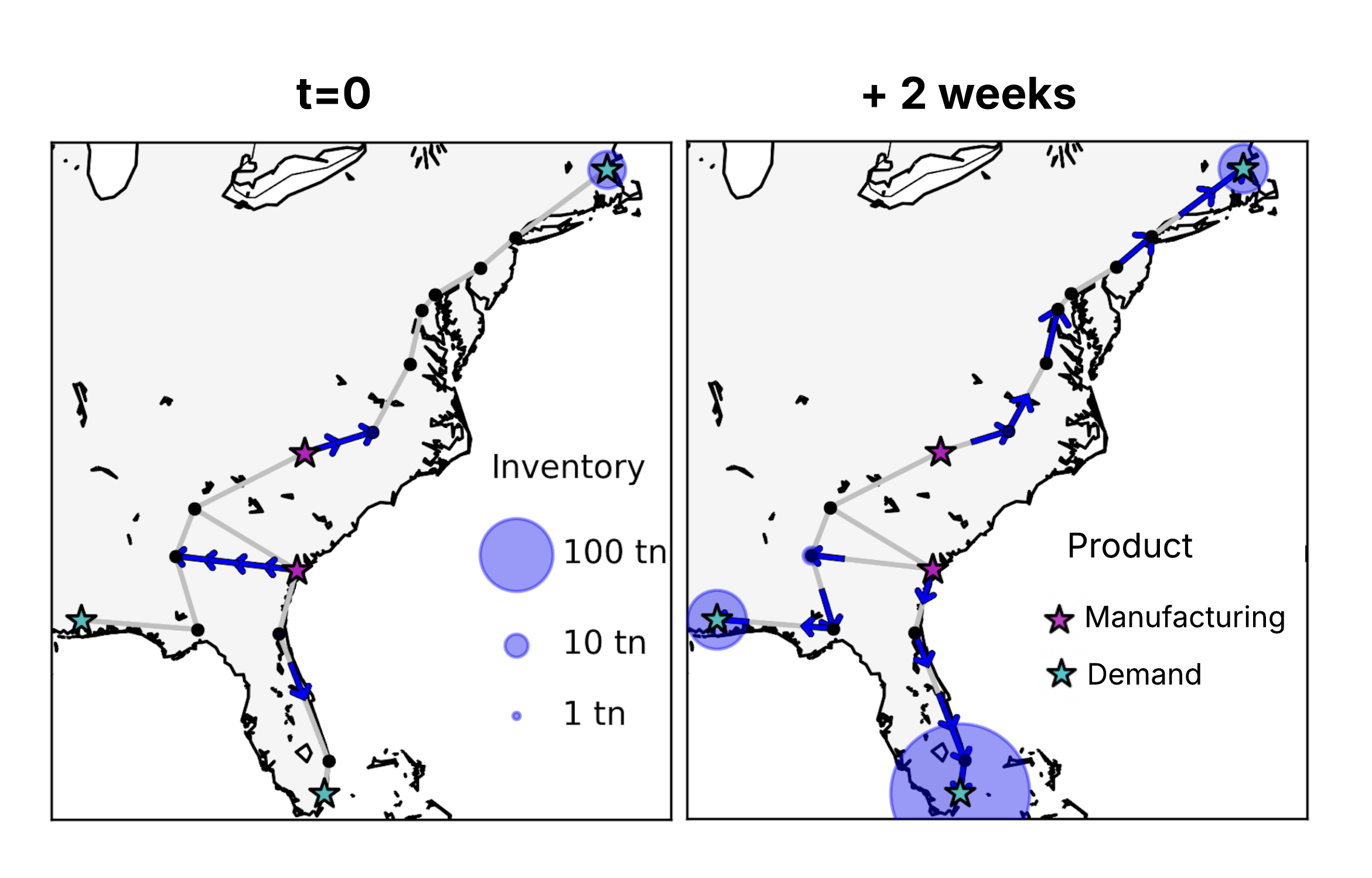}
    \caption{A visualization of the East Coast case study. Manufacturing locations are marked with pink stars, while demand locations are shown with blue stars. Arrows demonstrate the moving trucks, and the bubbles show the stockpiled inventory at each location. The right-hand side shows how the system evolves after two weeks of operation.}
    \label{fig:enter-label1}
\end{figure}

Our study uses a geographic network comprising 20 cities along the East Coast of the United States, ranging from Miami in the south to Boston in the north, and extending westward to Mobile. The cities are connected by links, which are each short enough for a truck to complete without stopping to charge. Two central manufacturing facilities with identical process parameters were placed in Charlotte, NC, and Savannah, GA. Fig. \ref{fig:enter-label1} visualizes this geographic network under the baseline scenario in which the supply chain is electrified, but there is no price differential between renewable and non-renewable power. We consider the hourly operation of the supply chain over a period of four weeks, during which each city's cement demand must be met by the end of the month.

To determine the renewable power available during each period of our case study, we used historical data from 2019 \cite{renewablesninja}, which provides hourly capacity factors for solar photovoltaic and wind power availability in each city. We assumed a one-to-one ratio of nameplate wind power to solar power. The total capacity at each city was 276 MW.

For some analysis presented later, we consider a reduced geographic network for computational simplicity. The smaller system comprises seven of the cities in the Southeast, and is discussed in greater detail in our previous publication~\cite{10741677}. 

\subsection{Simulation framework}
The case study used Python version 3.10.10 on a Dell XPS 16 Laptop containing the 13th Gen Intel Core i7-137000H and 16.0 GB of RAM. For longer time intervals where more memory was needed, we ran the problem using the PACE Phoenix cluster at Georgia Institute of Technology. The python packages CVXPY, NumPy, and Pandas were used, along with the optimization software Gurobi, which was utilized under an academic license to find a solution for the linear program.

\subsection{Reducing Complexity}
The formulation introduced above is a mixed-integer linear program (MILP), where several variables represent inherently integer qualities. Specifically, the variables which correspond to trucks: the number of full and empty trucks traversing each path $y,y'$, the number of stationary trucks $Y$, and the total fleet size $\hat{Y}$. However, solving as an MILP is computationally challenging. To achieve computational tractability, we relax the integer variables, which allows us to solve the problem faster. In our case, this was particularly important as we increased the size of our problem by lengthening the planning horizon and the number of locations available. 
The resulting solution gives an upper bound to the integer solution of the problem. 

\section{Results}

The four-week simulation was run for three scenarios: no carbon tax, \$50 carbon tax, and \$250 carbon tax. We see that with the introduction of the carbon tax, the operation of the supply chain changes noticeably -- highlighted in the following animation: \href{www.youtube.com/watch?v=tx5VAxwHEjg}{www.youtube.com/watch?v=tx5VAxwHEjg}. 

\begin{figure}[htbp]
    \centering
    \includegraphics[width=\columnwidth,trim={0 0.4cm 0 0.4cm},clip]{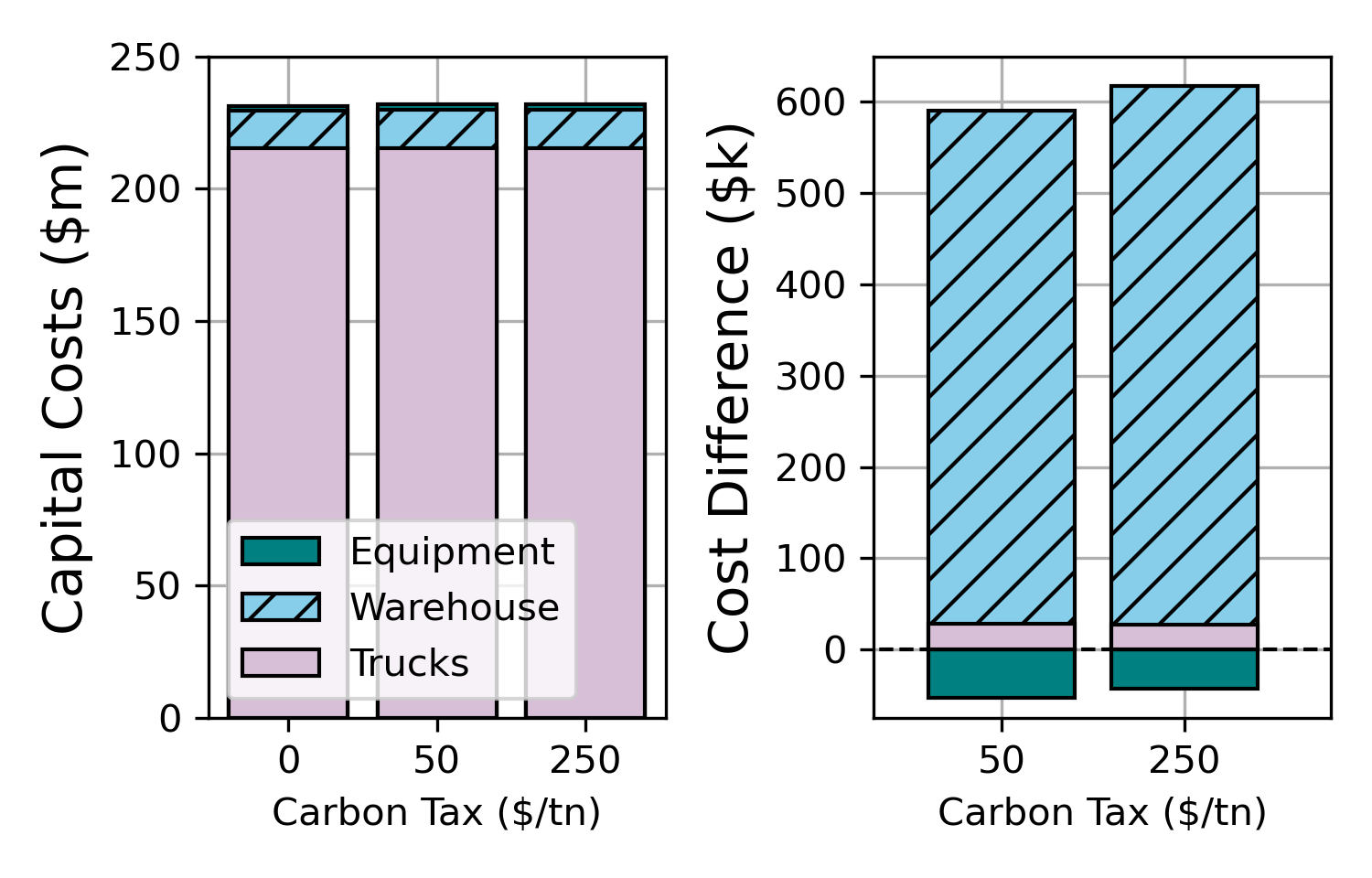}
    \caption{Illustrates the effect that different levels of carbon tax have on the supply chain capital expenditure, broken down into the equipment, trucks, and warehouse cost. }
    \label{fig:enter-label4}
\end{figure}

\begin{figure}[htbp]
    \centering
    \includegraphics[width=\columnwidth,trim={0 0 0 0},clip]{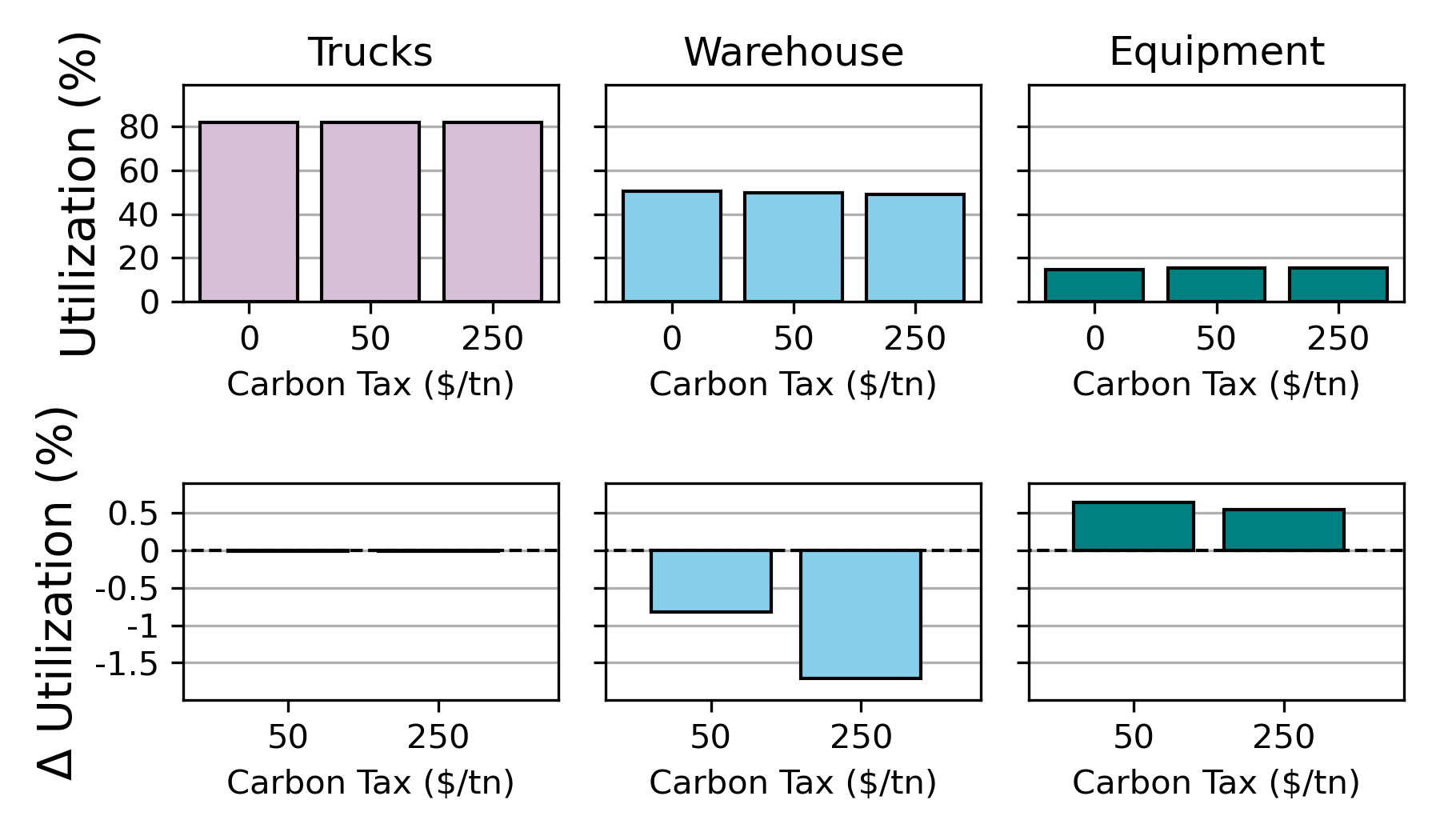}
    \caption{The average utilization percentages of trucks, warehouse, and equipment at different levels of a carbon tax (in \$/tn). The top plots show absolute utilization, and the bottom ones show the change from the zero carbon tax case.}
    \label{fig:enter-label2}
\end{figure}

Figure \ref{fig:enter-label4} shows how the optimal capital cost changes under varying levels of carbon tax, and Fig. \ref{fig:enter-label2} shows the changes in asset utilization. It is clear that the trucks dominate the capital expenditure in all cases. As the carbon tax increases, the main change we see is an increase in warehouse capacity and a slight reduction in manufacturing equipment. In all cases, the truck utilization is around 80\%, which is the maximum utilization given the time required for trucks to charge. Warehouse utilization decreases, which should be expected given the larger capacity. These results show that the primary change in response to the carbon tax is to the manufacturing schedule, meaning that more warehouse space is required to hold stockpiled finished products.

\begin{figure}[htbp]
    \centering
    \includegraphics[width=0.95\columnwidth,trim={0.35cm 0.45cm 0.3cm 0.1cm},clip]{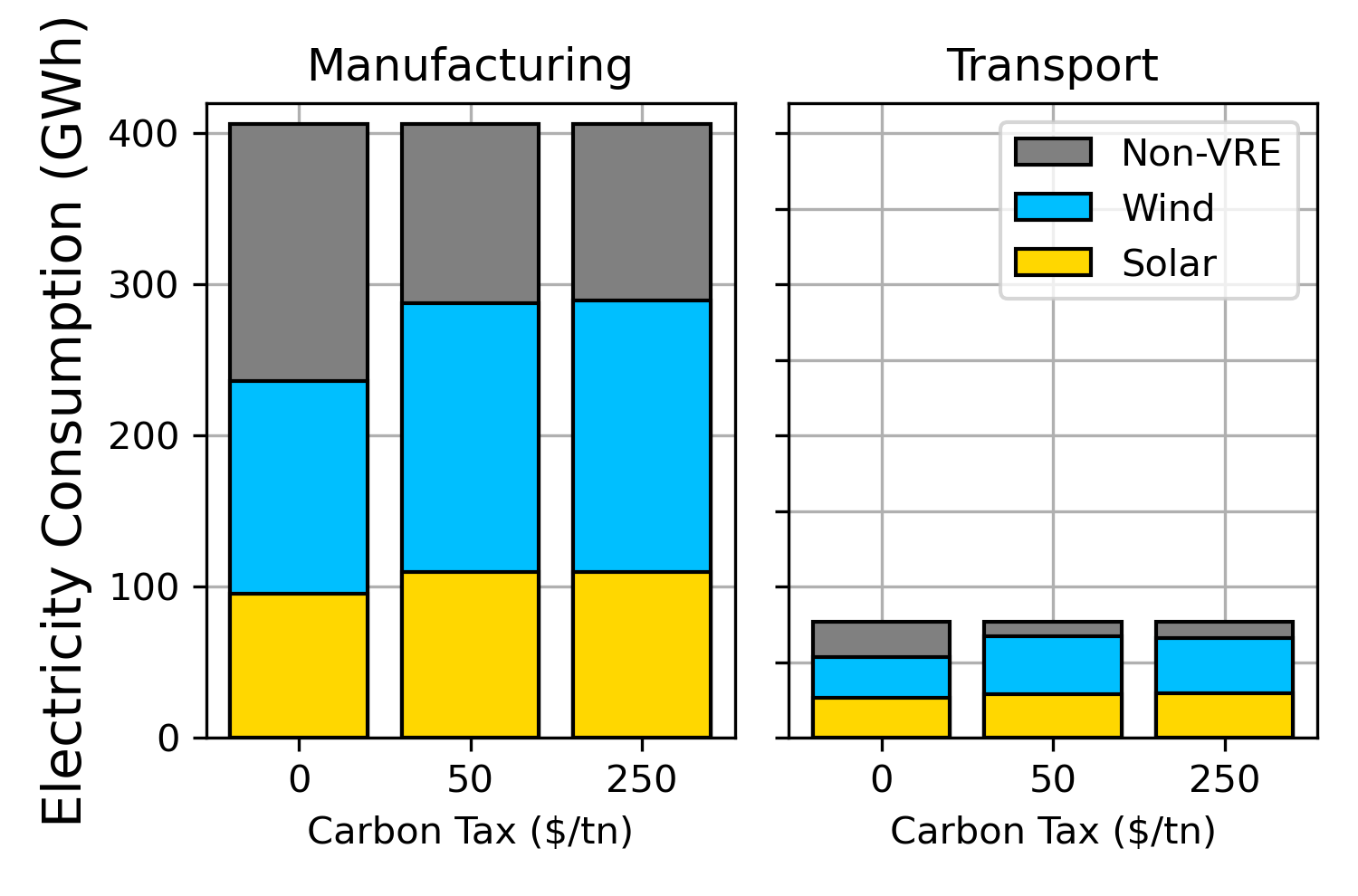}
    \caption{The total electricity consumption across all periods (in GWh) for manufacturing and transportation at different levels of carbon tax. The electricity consumed is broken down into non-renewable, wind, and solar energy. }
    \label{fig:enter-label3}
\end{figure}

Figure \ref{fig:enter-label3} shows the electricity consumption in each case, broken down by generation source. The left graph shows the energy used in product manufacturing, and the right shows the energy used for truck charging. It can be seen that manufacturing dominates energy consumption, due to the high energy demands of cement heating. It can be seen that there is a significant increase in renewables used. This increase is visibly dominated by additional wind energy consumption, potentially because wind power typically varies over longer timescales compared to solar, which has diurnal trends. Although we see an increase in renewables used by both the transport and manufacturing sectors, we know from Fig. \ref{fig:enter-label2} that the truck charging can not be significantly altered. Therefore, we can surmise that the manufacturing scheduling is adapted to prioritize renewables for truck charging.

\begin{figure}[htbp]
    \centering
    \includegraphics[width=\columnwidth,trim={0.4cm 0.4cm 0 0},clip]{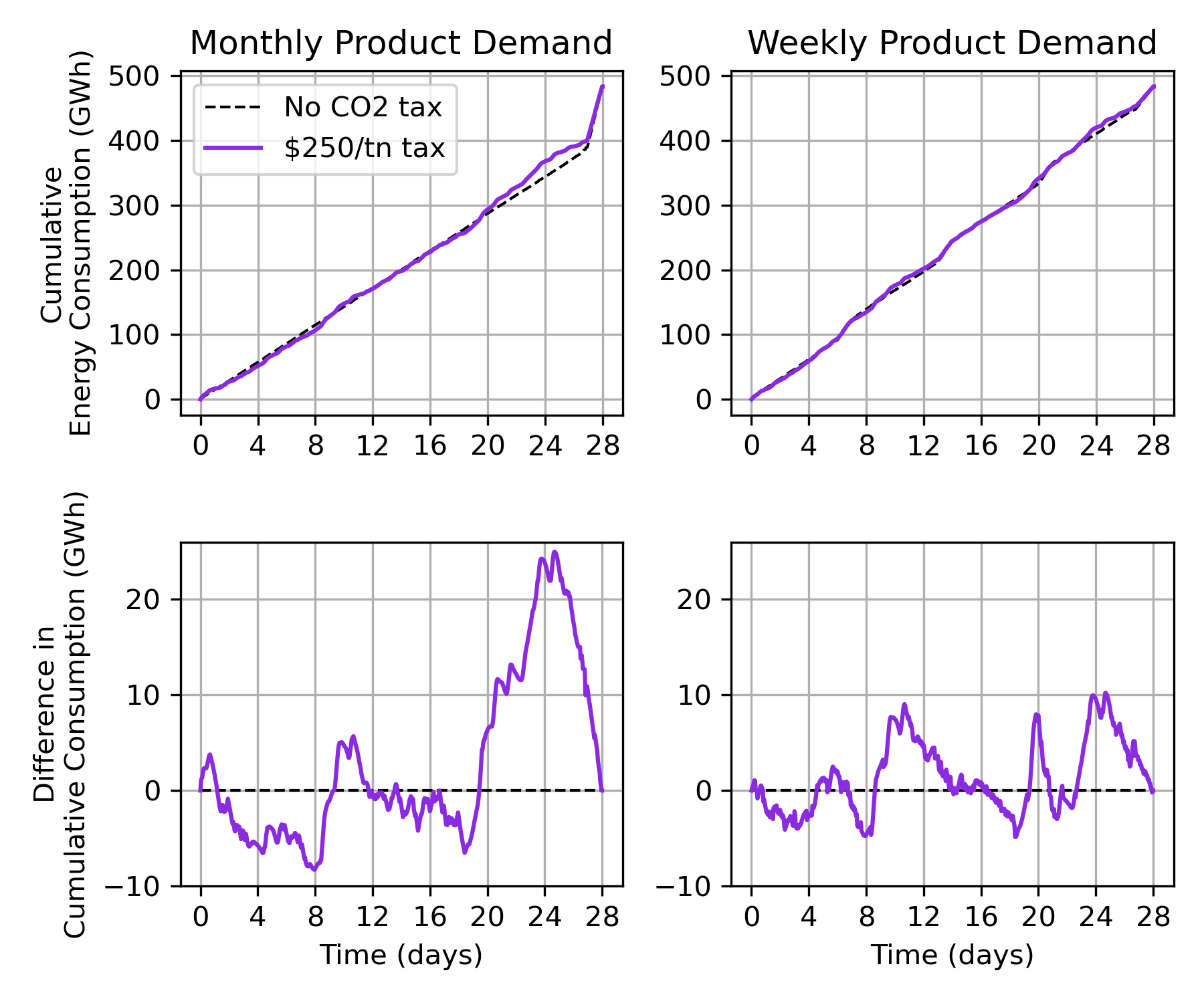}
    \caption{The cumulative energy consumption (GWh) throughout the simulation, and the difference in cumulative consumption (GWh) between weekly and monthly demand, under two carbon tax levels (\$0/tn and \$250/tn). A positive value implies earlier consumption, while a negative value is a delay.}
    \label{fig:enter-label5}
\end{figure}

Figure \ref{fig:enter-label5} analyzes the durations over which electricity demand is shifted. We compare the cumulative energy consumption over time with and without a carbon tax. Deviations from the baseline represent shifting demand, we expect that there will be zero deviation at the end of the simulation, given the finished product requirements. In addition to the case where the finished product demand must be cleared by the end of the month (monthly demand), we also consider the case where demand must be cleared each week (weekly demand). The latter represents a less flexible supply chain, and the difference in energy consumption must zero out at the end of each week instead of the end of the month. A positive difference signifies that demand is shifted forward (e.g., earlier manufacturing) while a negative difference signifies delayed manufacturing.

We see that for the case with monthly finished product clearing (monthly demand), the carbon tax incentivizes demand shifts of 20 GWh for over a week. This demonstrates the potential of supply chains for longer duration flexibility; existing shorter-duration demand flexibility must return to zero difference after hours. These longer variations allow the supply chain to better utilize wind power -- given that wind tends to vary over longer, less predictable timescales compared to solar. As expected, less significant volumes of demand are shifted in the weekly case, although in both cases, production is adjusted both forward and backward. 

Table \ref{t:summary} summarizes the energy and renewables used for the two cases of weekly and monthly demand at varying levels of carbon tax. We see that the amount of renewables used increases with the introduction of the carbon tax, but that \$250 achieves a negligible improvement over a \$50 carbon tax. The total energy demand is almost the same in all cases. This is because the manufacturing energy consumption is dominant, and the total energy consumed is directly related to the total finished product volume (which does not change). However, we do see small changes in the total trip distance; more trips are required when demand is cleared weekly, and in all cases, there is a slight reduction in distance traveled as the carbon tax increases. The total energy use from the supply chain, therefore, decreases only slightly, with the change on the order of kilowatt-hours relative to the gigawatt-hour total. 

We see no significant difference between the total renewables used in the \$50 and \$250 carbon tax cases. However, it should be noted that the vast majority of renewables available are being used. This suggests that the cheaper changes to the supply chain operation have already been saturated, and that there is insufficient financial incentive to motivate the changes necessary to use the last 0.1\% of available renewables. We observed the same effect in a number of different renewable scenarios, but there may well be some mix of renewable generation that would induce more distinction between \$50 and \$250. 

\begin{table}
\centering
\caption{Summary of the East Coast cement case study with weekly and monthly finished product demand}
\begin{threeparttable}
\small
\begin{tabular}{cccccc}
\toprule
\textbf{Carbon}     & \textbf{Product}  & \textbf{Distance}  & \textbf{Energy} & \multicolumn{2}{c}{\textbf{Renewables (\%)}}    \\ 
\textbf{Tax(\$/tn)} & \textbf{Demand} & \textbf{($10^3$ km)} & \textbf{(GWh)} & \textbf{Demand} & \textbf{Avail} \\  
\midrule
\$0   & Weekly  & 39581.1   & 483   & 59.53 & 80.97 \\  
\$50  & Weekly  & 39580.0   & 483   & 73.46 &  99.91\\  
\$250 & Weekly  & 39579.8  & 483  & 73.46 & 99.91\\  
\midrule
\$0   & Monthly & 39579.7  & 483  & 59.86 & 81.42\\ 
\$50  & Monthly & 39579.4   & 483  & 73.46 & 99.91 \\  
\$250 & Monthly & 39579.3 & 483  & 73.46 & 99.91\\  
\bottomrule 
\end{tabular}
\label{t:summary}
\end{threeparttable}
\end{table}

\subsection{Comparing to Battery Energy Storage}

We also present a comparison between leveraging supply chain flexibility and installing battery energy storage systems (BESS). The intent is to compare the cost-effectiveness of using supply chain flexibility compared to grid-scale energy storage. The main takeaway from Table \ref{t:Combined} is that using supply chain flexibility resulted in a significantly higher percentage of renewables compared to BESS, under a substantially lower carbon tax. We see that at low carbon taxes, no batteries are installed, meaning it becomes economic to utilize supply chain flexibility long before it becomes economic to install grid storage. As the carbon tax increases, additional battery capacity is installed, but still in very small amounts. At \$400/tn the installed battery capacity is still smaller than the battery of a single truck. The results also demonstrate that the supply chain flexibility is actually incentivized at much lower carbon taxes than the \$50/tn we previously considered -- in this case, the same benefit is achieved at \$1/tn or \$0.39/MWh. 


\begin{table}[h!]
\centering
\caption{Renewable Energy Utilization for the Proposed Model (Supply Chain Flexibility) and BESS under Varying Carbon Tax Levels}
\begin{tabular}{c|c|cc}
\toprule
\textbf{Carbon Tax} & \textbf{Max Battery Capacity} & \multicolumn{2}{c}{\textbf{Renewable Percentage (\%)}} \\ 
\textbf{(\$/tn)} & \textbf{(kWh)} & \textbf{Proposed} & \textbf{BESS} \\ \midrule
0   & 0.0    & 38.23 & 38.23 \\
0.01  & 0.0  & 61.73  & 38.23 \\
1  & 0.0    & 74.26  & 38.23 \\
50  & 0.0    & 74.26 & 38.23 \\
100 & 0.0   & 74.26 & 38.23 \\
200 & 18.71 & 74.26 & 39.45 \\
250 & 58.83 & 74.26 & 41.69 \\
300 & 98.54 & 74.26 & 43.53 \\
350 & 121.8 & 74.26 & 44.42 \\
400 & 145.78 & 74.26 & 45.23 \\
\bottomrule
\end{tabular}
\label{t:Combined}
\end{table}
\section{Sensitivity Analysis}

\begin{figure}[htbp]
    \centering
    \includegraphics[width=\columnwidth,clip,trim={0 3cm 4cm 4cm}]{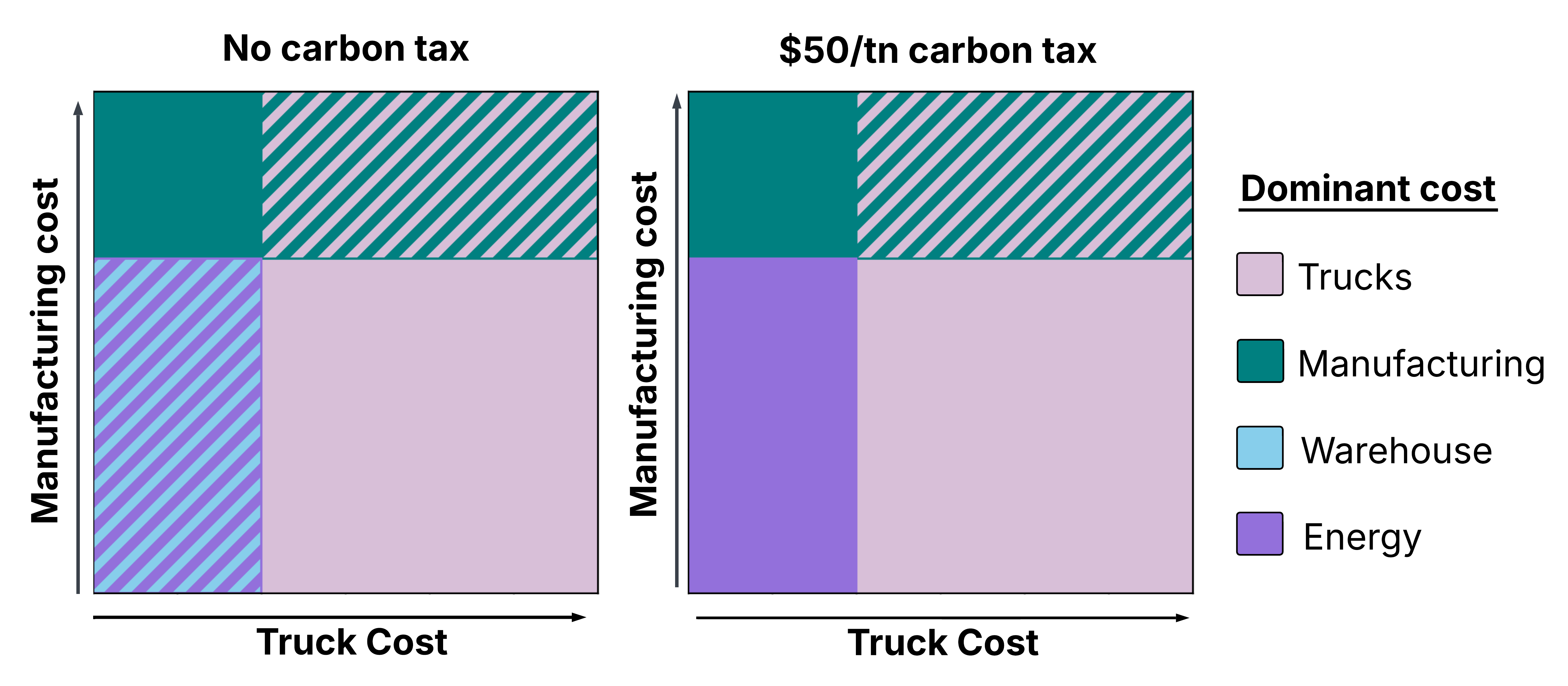}
    \caption{A visualization of how the supply chain adapts to minimize the quantities of trucks, manufacturing, warehousing, and energy consumption as truck and manufacturing costs vary. Two carbon tax levels are considered: no carbon tax and \$50/ton. The x-axis shows the impact of increasing truck costs, while the y-axis shows the impact of increasing manufacturing costs. }
    \vspace{-.2cm}
    \label{fig:sensitivity}
\end{figure}

The results presented in the previous section are specific to our case study of electrified cement manufacturing on the East Coast. However, to understand how these results might be generalized to other supply chains, we conducted a sensitivity analysis on the capital costs. Here, we focus on the truck cost and manufacturing equipment costs, as these have the most significant effects. Note that changes in cost can also be a proxy for different parameters of the finished product, for example, lower truck cost could replace a denser product with a relatively smaller transportation need. 
We analyzed a series of combinations in which truck and manufacturing costs were set to 0.1 times, 10 times, or equal to their original prices. The results are summarized in Table \ref{t:sensitivity} and Fig. \ref{fig:sensitivity}.

\begin{table*}[!t]
\centering
\renewcommand{\arraystretch}{1.2}
\caption{Sensitivity Analysis of Results with Varying Truck and Manufacturing Costs}
\begin{threeparttable}
\begin{tabular}{|c|c|cc|cc|cc|cc|cc|cc|}
\hline
\multirow{4}{*}{\textbf{\begin{tabular}[c]{@{}c@{}}Truck Cost\\ Scale \\ Factor\end{tabular}}} & \multirow{4}{*}{\textbf{\begin{tabular}[c]{@{}c@{}}Manufacturing \\  Cost Scale \\ Factor\end{tabular}}} & \multicolumn{2}{c|}{\textbf{Energy}} & \multicolumn{2}{c|}{\textbf{Renewables}} & \multicolumn{2}{c|}{\textbf{Truck Fleet}} & \multicolumn{2}{c|}{\textbf{Manufacturing}} & \multicolumn{2}{c|}{\textbf{Warehouse}} & \multicolumn{2}{c|}{\textbf{Distance}} \\
 & & \multicolumn{2}{c|}{\textbf{(GWh)}} & \multicolumn{2}{c|}{\textbf{(\%)}} & \multicolumn{2}{c|}{\textbf{(\# Vehicles)}} & \multicolumn{2}{c|}{\textbf{(Units)}} & \multicolumn{2}{c|}{\textbf{($\mathbf{10^6}$kg)}} & \multicolumn{2}{c|}{\textbf{($\mathbf{10^6}$km)}} \\[.1cm] 
\cline{3-14}
 && \multicolumn{2}{c|}{Carbon Tax} & \multicolumn{2}{c|}{Carbon Tax} & \multicolumn{2}{c|}{Carbon Tax} & \multicolumn{2}{c|}{Carbon Tax} & \multicolumn{2}{c|}{Carbon Tax} & \multicolumn{2}{c|}{Carbon Tax} \\
& & \$0 & \$50  & \$0 & \$50  & \$0 & \$50  & \$0 & \$50  & \$0 & \$50 & \$0 & \$50 \\  
\midrule
0.1 & 0.1 & \textcolor{white}{x}309\textcolor{white}{x}  & \textcolor{white}{x}307\textcolor{white}{x} & \textcolor{white}{x}4.45\textcolor{white}{x} & \textcolor{white}{x}76.5\textcolor{white}{x}  & 19,900 & 28,400 & \textcolor{white}{x}225\textcolor{white}{x} & 58.1 & \textcolor{white}{x}63.1\textcolor{white}{x} & 553 & 4.87 & 1.71 \\
0.1 & 1   & 309 & 307 & 4.45 & 76.5 & 19,900 & 28,400 & 220 & 58.1 & 68.3 & 553 & \textcolor{white}{x}4.87\textcolor{white}{x} & \textcolor{white}{x}1.71\textcolor{white}{x}  \\
0.1 & 10  & 309 & 307 & 41.8 & 76.5 & 19,900 & 28,400 & 8.08$^\dagger$ & 8.08 & 776 & 776 & 4.87 & 1.71 \\
1   & 0.1 & 316 & 316 & 46.5 & 74.3 & 476$^\dagger$  & 476$^\dagger$ & 95.2 & 53.7 & 544 & 615 & 12.5 & 12.5 \\
1   & 1   & 316 & 316 & 38.2 & 74.3 & 476$^\dagger$ & 476$^\dagger$ & 75.3 & 53.7 & 549 & 615 & 12.5 & 12.5 \\
1   & 10  & 316 & 316 & 58.6 & 74.3 & 476$^\dagger$ & 476$^\dagger$ & 8.08$^\dagger$ & 22.4 & 784  & 776 & 12.5 & 12.5 \\
10  & 0.1 & 316 & 316 & 45.8 & 74.3 & 476$^\dagger$  & 476$^\dagger$ & 95.2 & 53.7 & 544 & 615 & 12.5 & 12.5 \\
10  & 1   & 316 & 316 & 36.4 & 74.3 & 476$^\dagger$ & 476$^\dagger$ & 75.3 & 53.7 & 549 & 615 & 12.5 & 12.5 \\
10  & 10  & 316 & 316 & 58.6 & 74.3 & 476$^\dagger$  & 476$^\dagger$ & 8.08$^\dagger$ & 22.4 & 784 & 728 & 12.5 & 12.5 \\
\bottomrule
\end{tabular}
\label{t:sensitivity}
\begin{tablenotes}
\item[$\dagger$] Indicates that the value is the lowest feasible that still meets the system's basic requirements (in other words, the value obtained by explicitly minimizing, e.g., the number of trucks).
\end{tablenotes}
\end{threeparttable}
\end{table*}

For all cost combinations, there is a substantial increase in renewable utilization once a carbon tax is introduced. This is especially notable given the differences across the various supply chain configurations considered. The highest sensitivity is to truck costs; in most cases, the operation of the supply chain revolves around minimizing the required number of trucks. However, for some cost combinations we see that the supply chain can pivot to prioritize: minimizing manufacturing equipment, minimizing warehouse space, or minimizing energy costs. For example, when trucks are ten times cheaper and manufacturing equipment is ten times as expensive, the supply chain changes to ensure the continual operation of the manufacturing equipment, with additional warehouses and trucks purchased to supply energy and transport finished products whenever needed. 

In each case, the objective function becomes dominated by one or two of the cost terms, and the remaining components are sized to ensure those dominant terms are minimized. There is inherent latency in the other components, which can be utilized at low cost for power grid flexibility. It is likely that many supply chains will have components that dominate costs around which the remainder of the system will be designed, thus possessing latency and the potential to provide grid flexibility.

\section{Discussion \& Conclusions}
This paper demonstrated that supply chains may be able to provide demand flexibility for durations of over a week at minimal cost. This arises due to one or two components dominating the capital costs, meaning the supply chains often possess latent flexibility in the other components, which can be leveraged to increase the utilization of variable renewable energy. To explore this potential, a linear programming model was developed to manage the operation of a supply chain with electrified manufacturing and transportation.

To estimate the impacts of leveraging supply chain flexibility, we developed a simplified case study involving 20 cities along the East Coast of the U.S., with cement as the sole finished product being produced and distributed. We assessed the effect of a carbon tax on non-renewable power (\$50/ton equivalent to $~\$19$/MWh) on the amount of flexibility provided. Overall, the case study demonstrates that longer-duration demand shifts are possible, even at very low levels of carbon tax (as low as \$1/tn). We demonstrated that the shift in demand is motivated by variation in wind power, which varies over longer durations than solar power. 

We demonstrated that flexibility is created by increased warehouse and manufacturing capacity, allowing the manufacturing and delivery schedule to be altered. The utilization of electric trucks remains constant, as the truck capital costs dominate, and thus the system acts to minimize the required number of trucks. We observed greater flexibility in energy consumption when the product demand must be met by the end of the month, compared to the case where demand is cleared weekly. 

A sensitivity analysis was conducted on a smaller network consisting of a subset of seven cities. In the analysis, we demonstrated that most supply chains will be designed around dominant costs (e.g. the cost of trucks). The remainder of the system is designed around maximum utilization of these assets, resulting in latent flexibility. However, as the costs change, different costs can become dominant, and the system can pivot to prioritize another cost (e.g. manufacturing).

A comparison with battery energy storage systems showed that using supply chain demand flexibility achieved higher renewable utilization at far lower costs compared to installing battery storage. 

\subsection{Limitations and Future Work}

The work presented in this paper made several simplifying assumptions. The results shown use continuous relaxation of the integer variables and neglect the modeling of the workforce or multi-stage manufacturing problems. Here we describe several further directions that will allow more accurate estimates of flexibility potential to be made.

A key limitation of the supply chain flexibility model developed here is its computational complexity. Even using HPC resources, we were forced to relax integer variables and could only consider time-horizons of four weeks. The four week period limits the flexibility potential of our simulation, as products can only be stockpiled over weeks. Meanwhile, the continuous relaxation leads to an overestimate of the flexibility potential. Decomposition methods may allow us to find integer solutions to larger and more accurate case studies, which would provide more reliable estimates of long-duration flexibility potential. 

Our current model considers a single-stage manufacturing process. Future work could explore multiple manufacturing facilities, with multiple processes that feed each other. Significant complexity could also be added to the power system component of our model. For example, future analysis could consider the placement of renewable generation and charger facilities (rather than using fixed solar and wind capacities). This analysis would require the inclusion of power grid constraints, rather than assuming that electricity can be freely transmitted between locations. 

Finally, our supply chain model assumes deterministic values for: renewable energy availability, the supply of raw materials, imports of additional finished products, and demand for the finished product. In future work, the stochastic nature of these parameters can be incorporated, which would also require accounting for the costs and revenues associated with imports and sales, as they will no longer be fixed.

\section*{Appendix}
Here we detail our derivation of the parameters for the cement manufacturing process. 
The assumptions listed in Table \ref{t: Parameters} were used to model the key characteristics of cement manufacturing in the most realistic manner possible. These assumptions, including warehouse dimensions, cement properties, and kiln production rate, were crucial for estimating process parameters. We assume use of the primary technology that electrifies cement, Coolbrook's Roto Dynamic Heater \cite{coolbrook_rdh}, capable of reaching temperatures up to 1700$^\circ C$.

\begin{table}[h!]
\centering
\caption{Assumptions for cement case study}
\begin{tabular}{l|l|l}
\toprule
\textbf{Parameter} & \textbf{Value} & \textbf{Unit} \\
\midrule
Warehouse Size & 4,645 & m\textsuperscript{2} \\
Warehouse Lifetime & 30 & year \\
Warehouse Height & 9.88 & m \\
Warehouse Construction Cost (4,645 m\textsuperscript{2}) & 1,150,000 & USD \\
Cement Density & 1,440 & kg/m\textsuperscript{3} \\
Cement Kiln Production Rate & 150,000 & kg/hour \\
Cement Specific Heat Capacity ($c$) & 0.92 & kJ/kg\textdegree C \\
Cement Temperature Change ($\Delta T$) & 1,400 & \textdegree C \\
Electric Cement Heater Cost & 25,000 & USD \\
\bottomrule
\end{tabular}
\label{t: Parameters}
\end{table}

The warehouse volume was calculated using an average warehouse size and an estimated warehouse height. Using the density of cement, the storage capacity per-square meter was determined. Construction costs were adjusted on a per-square meter basis, and the cost per kilogram of cement was levelized over the warehouse's lifetime. \\
\indent Energy for the kiln was estimated using the formula
\[
E = mc\Delta T,
\]
where \( m \) is the kiln production rate, \( c \) is the specific heat capacity, and \( \Delta T \) is the temperature change. The energy was then converted to kWh. The capital cost of the kiln was assumed to be twice that of a traditional kiln, \$12,500. The kiln's lifetime was estimated at 40 years, and the levelized cost was calculated accordingly.


\section*{Acknowledgment}
We would like to recognize the contributions of our collaborators: Chelsea White from the Georgia Institute of Technology, as well as Eman Hammad, Elnaz Kabir, Eleftherios Iakovou, and Halil Iseri from Texas A\&M University.

\bibliography{refs.bib}{}
\bibliographystyle{IEEEtran}

\end{document}